# Design and Implementation of Ultra-Wideband Dual-Polarized Conformal Phased Array

Zhan Chen, Wei Hu, *Senior Member, IEEE*, Yuchen Gao, Xiangbo Wang, Qi Luo, *Senior Member, IEEE*

***Abstract*—This letter presents an innovative design and implementation method for an ultra-wideband dual-polarized conformal phased array. The performance of single-layer continuous dipoles under bending conditions is evaluated through characteristic mode analysis to understand the impact of curvature on operational stability. This analysis provides design insights into the maximum allowable curvature while maintaining stable dipole performance. Subsequently, a single-layer metal radiation structure is designed as the array element, and simulations indicate that it supports ultra-wideband operation with dual polarizations. Using this element, an 8 × 8 ultra-wideband dual-polarized cylindrical-conformal array (UDCA) is developed, demonstrating stable performance even at a curvature radius as small as 100 mm. A physical prototype is fabricated cost-effectively using novel manufacturing techniques that stack a three-layer conformal substrate. Experimental result demonstrates that the proposed UDCA with a 1.2λ curvature radius operates at 3.6~9.6 GHz (90.9%) and achieves ±60° wide-scanning in E-/H-planes, which provides a practical and promising solution for conformal array applications.**

***Index Terms*—Characteristic mode analysis, conformal array, dual-polarized, ultra-wideband, wide-angle scanning.**

## I. Introduction

Conformal antennas are favored for high-speed maneuverable platforms, which enhance the aerodynamic performance without altering the shape of aircraft [1]. Furthermore, the integration of more functionalities into large-aperture conformal antennas serves to diminish the count and intricacy of airborne antennas, enhancing the overall efficiency of antenna layout [2]. For such applications, conformal antennas are required to ultra-wideband and dual-polarized operation.

Various types of conformal antennas have been investigated including cylindrical, conical, spherical, and irregular surfaces, which operate within a narrow relative bandwidth of below 5% [3]-[13]. Conventional conformal antenna design for enhancing bandwidth can be divided into two main categories. One is to sacrifice the profile of the antennas [14]-[18]. For this type of conformal antenna, the high profile limits the application on carrier platforms, especially for large conformal curvature. Another method is to expand the lateral dimensions [19]-[21]. The dimensions of these conformal antenna units are greater than the half-wavelength of the lowest operating frequency, which is not suitable for the beam-scanning array applications.

Planar tightly coupled arrays have received significant attention due to their low-profile, ultra-wideband, and wide-angle scanning operation characteristics [22]-[27]. A novel characteristic excitation for tightly coupled arrays by using the characteristic mode analysis (CMA) provides wideband matching of all array elements [28]. However, complex tightly coupled arrays are challenging to design and manufacture for curved applications [29]. Due to the spacing constraints for feeding, a conformal array is proposed by using a dual-polarized unit cell to achieve single-polarized operation [30]. A single-polarized flexible tightly coupled conformal array with 9:1 operating bandwidth by loading frequency selective surface is reported without beam scanning capability [31]. More importantly, these antenna designs from planar infinite arrays to conformal finite arrays require extensive parameter optimization calculations which is time-consuming. Therefore, there is an urgent requirement to find an effective design method that can simplify the design process of the conformal array antenna and meet the targeted performance.

In this letter, a novel design and implementation method of ultra-wideband dual-polarized conformal phase array is presented. CMA for continuous dipoles gives the design of the conformal antenna a guideline for using a single-layer metal radiation structure. The developed ultra-wideband dual-polarized cylindrical-conformal array (UDCA) shows stable high operating performance under conformal bending. The physical prototype is cost-effectively fabricated that stacks a three-layer conformal substrate. This presented method offers a feasible and efficient solution for applications requiring large-curvature conformal antenna.

## II. CMA For Continuous Dipole

In the theory of CMA [32] [33], mode significance (MS) provides a convenient way to measure the operation performance of each CM. The relationship between the impedance, current, and radiation pattern corresponding to MS of each CM has been demonstrated [34] [35]. Meanwhile, the

This work was supported in part by the National Natural Science Foundation of China under Grant No. 62201417, and in part by the 111 Project of China.
(*Corresponding author: Wei Hu*)

Zhan Chen, Wei Hu, Yuchen Gao, and Xiangbo Wang are with the Nationa l Key Laboratory of Antennas and Microwave Technology, Xidian University, Xi'an, Shaanxi 710071, China (e-mail: weihu.xidian@ieee.org).

Qi Luo is with the School of Physics, Engineering and Computer Science, University of Hertfordshire, AL10 9AB Hatfield, U.K (e-mail: qiluo@ieee.org).

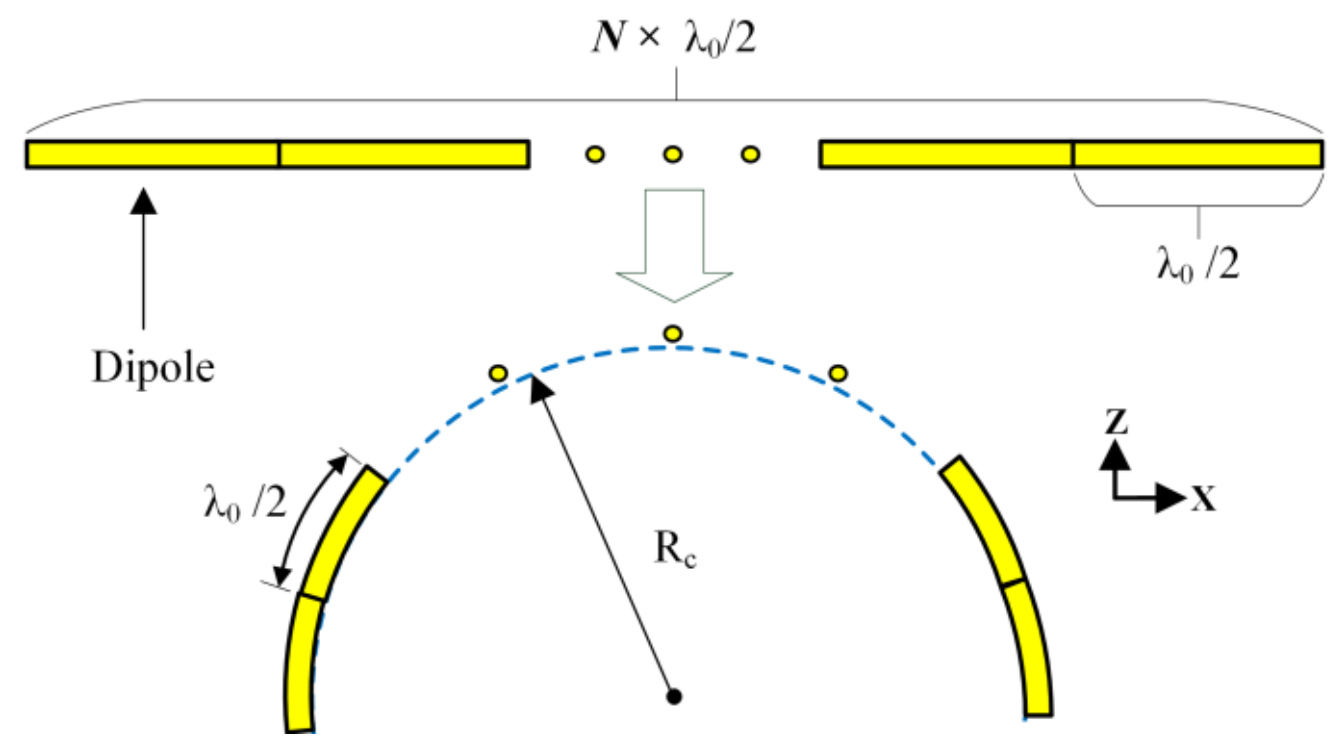

Fig. 1. The schematic diagram for continuous dipoles structure under bending. ($\lambda_0$ is the wavelength of 5 GHz in free space.)

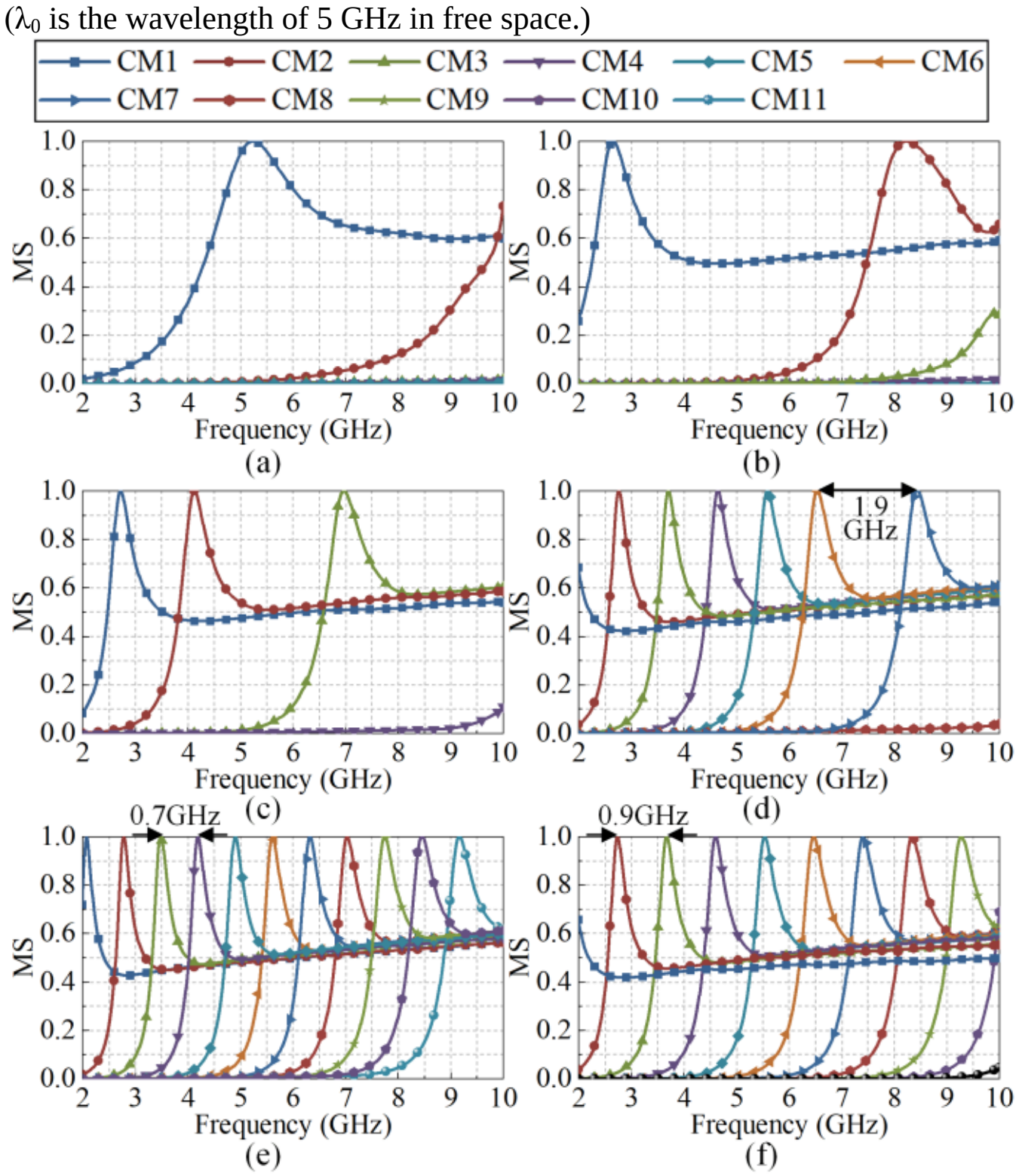

Fig. 2. The MSs for varying continuous dipoles. (a) $N$ = 4. (b) $N$ = 6. (c) $N$ = 8. () $N$ = 10.

CMA of antenna without feeding can be used to guide the antenna design with feeding, which improve the assessment efficiency of antenna performance during conformal bending of radiation structure.

The CM distribution for varying numbers of planar continuous half-wave dipoles is first investigated, as shown in Fig. 1. When $N$ = 1, the half-wave dipole operates at 5 GHz in half-wave mode ($CM_1$) as shown in Fig. 2(a). As $N$ increases, more CMs are generated to operate within the 2 to 10 GHz band as evidenced from Fig. 2(b) to 2(f). For $N$ = 8, the operated 11 CMs can cover the entire frequency band of 2.0~10.0 GHz, where the interval frequency for the MS = 1 of each two CMs averages 0.7 GHz. At $N$=6, the interval frequency between $CM_6$ and $CM_7$ reaches 1.9 GHz, whereas at $N$=10, the interval frequency of each two adjacent CMs reaches 0.9 GHz on average. Due to the small frequency separation, it is easier to achieve wideband coverage. As a result, the $N$ = 8 is chosen for the ultrawideband antenna design.

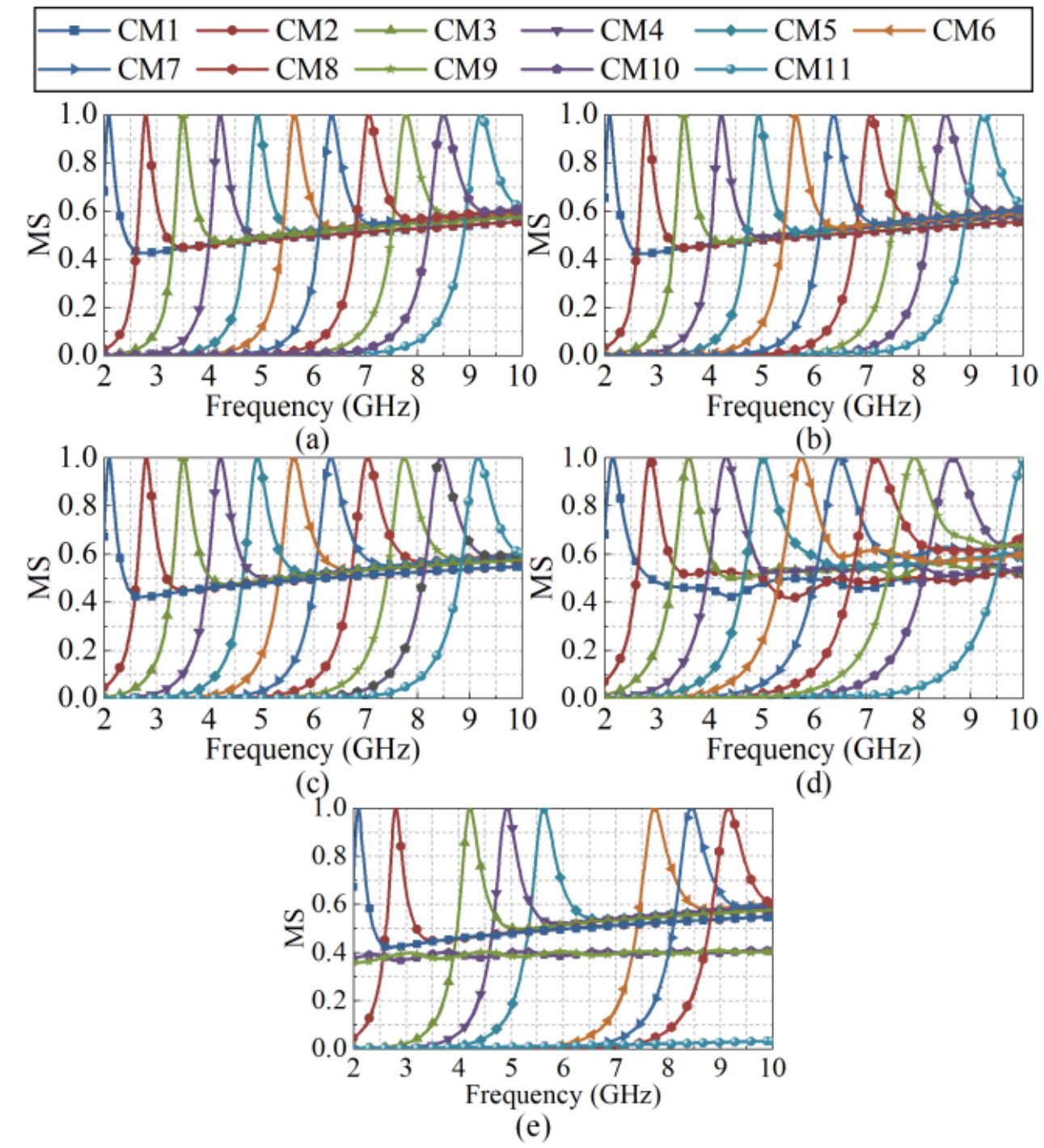

Fig. 3. The MSs for variable curvature radius with the continuous half-wave dipole at $N$ = 8. (a) $R_c$ =200 mm. (b) $R_c$ =150 mm. (c) $R_c$ =100 mm. (d) $R_c$ =50 mm. (e) $R_c$ =33 mm with forming a complete loop.

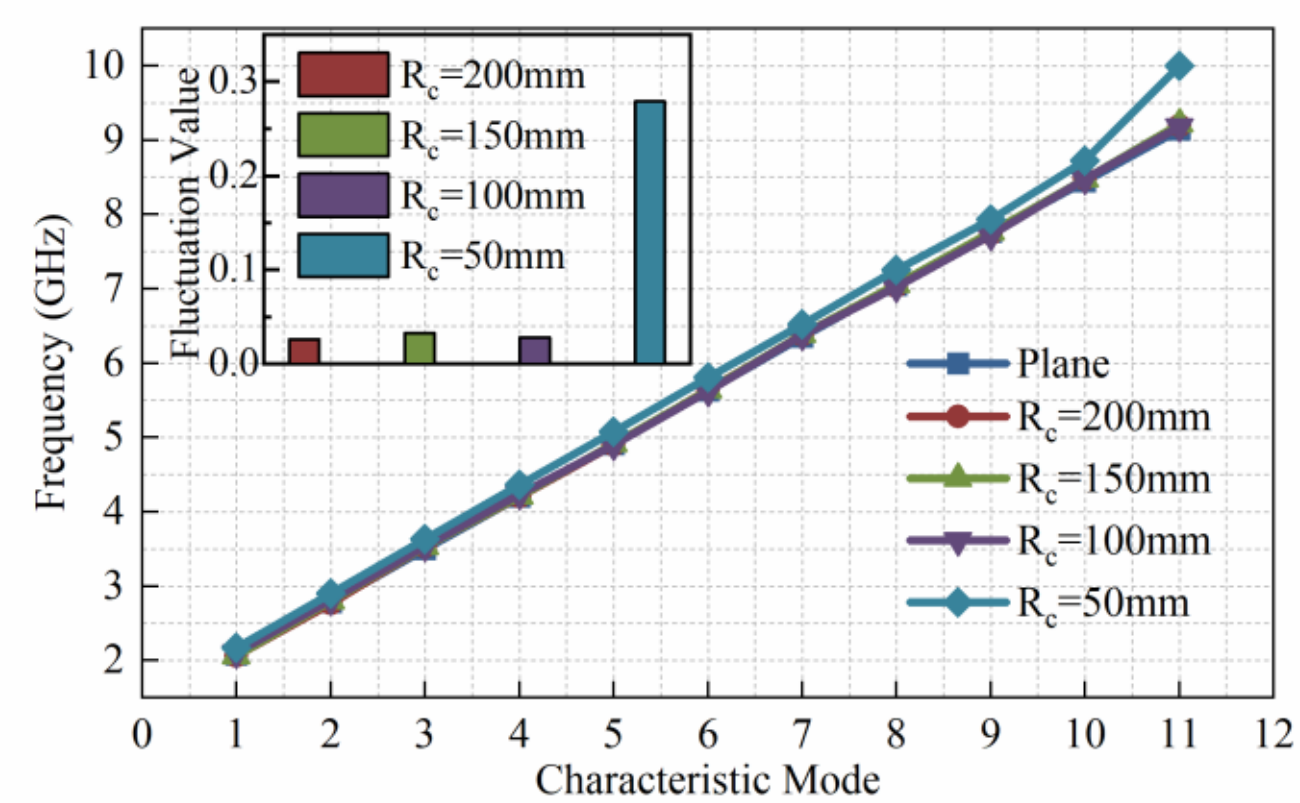

Fig. 4. The simulated frequency values at MS=1 for first 11 CMs in bending operation at $N$ = 8.

The next step is to investigate bending the continuous dipoles with $N$ = 8, where the MSs for variable curvature radius $R_c$ are illustrated in Fig. 3. For $R_c$ = 200 mm, 150 mm, and 100mm, the MS values of the given the first 11 CMs are almost identical. When reducing the $R_c$ to 50 mm, it is found that there are no corresponding CMs operated near the 9.0 GHz as $CM_{11}$ is shifted towards higher frequencies. Thus, at a radius of curvature of 50 mm, the operated performance of the higher-order modes of the dipole operating in a conformal design is altered. Further, the structure forms a loop when $R_c$ = 33 mm, at which the distribution of the CMs has completely changed, indicating that there is a significant abrupt variation in the operating performance.

The simulated frequency values of the first 11 CMs at MS=1 for bending operation are given in Fig. 4. In $R_c$ = ∞, 200 mm, 150 mm, and 100 mm, the four curves almost coincide, whereas when $R_c$ = 50 mm, the frequency shift of the CMs becomes

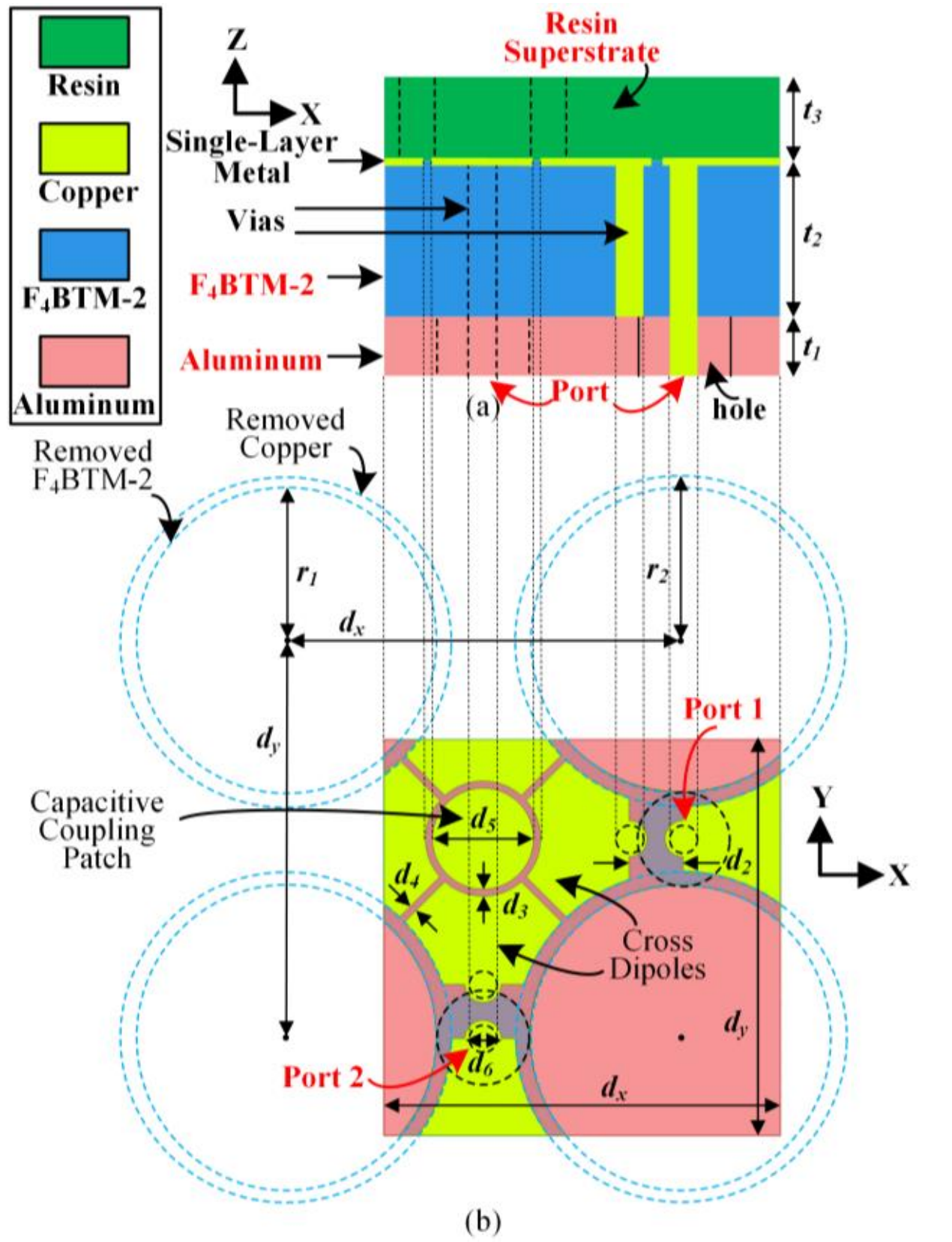

Fig. 5. Geometry of periodic unit cell for the dual-polarized ultra-wideband array. (a) main view. (b) Top view. $d_x = d_y = 11$, $d_1 = 0.8$, $d_2 = 1.5$, $d_3 = 0.2$, $d_4 = 0.2$, $d_5 = 2.8$, $d_6 = 1$, $r_1 = 4.2$, $r_2 = 4.6$, $t_1 = 1.8$, $t_2 = 4.75$, $t_3 = 3$. (All dimensions in mm)

more significant the higher the operating frequency. In order to more easily observe how these CMs change under dipole bending, the fluctuation value (FV) of frequency is defined as:

$$\mathrm{FV} = \frac{\sqrt{\sum_{i=1}^{n}\left(f_{\mathrm{CM}i}^{\mathrm{A}} - f_{\mathrm{CM}i}^{\mathrm{B}}\right)^2}}{n} \tag{1}$$

where A and B represent different curvatures, $f_{\mathrm{CM}i}^{\mathrm{A}}$ and $f_{\mathrm{CM}i}^{\mathrm{B}}$ is the frequency value of the $\mathrm{CM}_i$ for MS = 1 with curvature A and curvature B, respectively. The smaller the FV value, the less variation in the CMs is represented. The FV values are calculated for each radius of curvature using the FV of the planar structure as a reference value. At $R_c$=50mm, the FV is 0.279, While the other three curvature states have FVs less than 0.05. Thus, the continuous dipoles for the curvature radius $R_c$ within 100 mm can maintain the stability of its operating performance when the $N$ = 8, which became the optimized choice for the maximum curvature.

## III. Planar Infinite Array Design

After performing the analysis shown in the previous section, the antenna element that is suitable for an ultra-wideband conformal array should satisfy the following conditions.

1) The unit cells within the array should be tightly connected that obtain a wideband operation.

2) The radiation structure of the antenna should be single-layer to maintain stable performance in conformal bending.

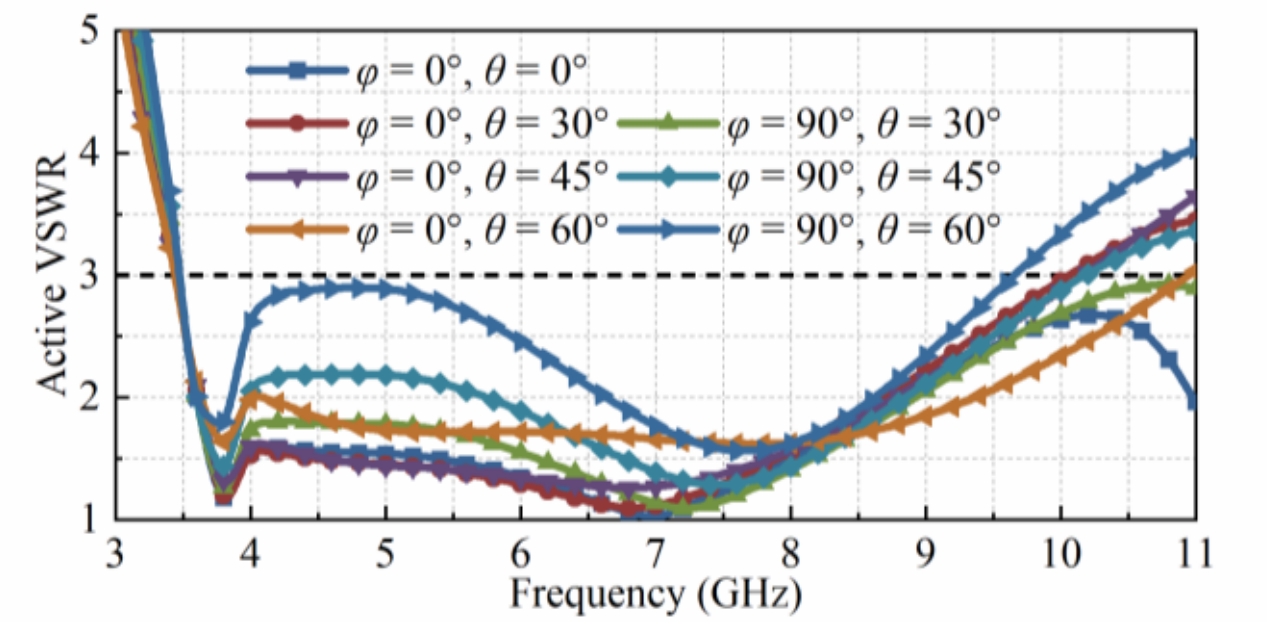

Fig. 6. Simulated active VSWRs of X-polarized unit cell for infinite array at various scanning angles.

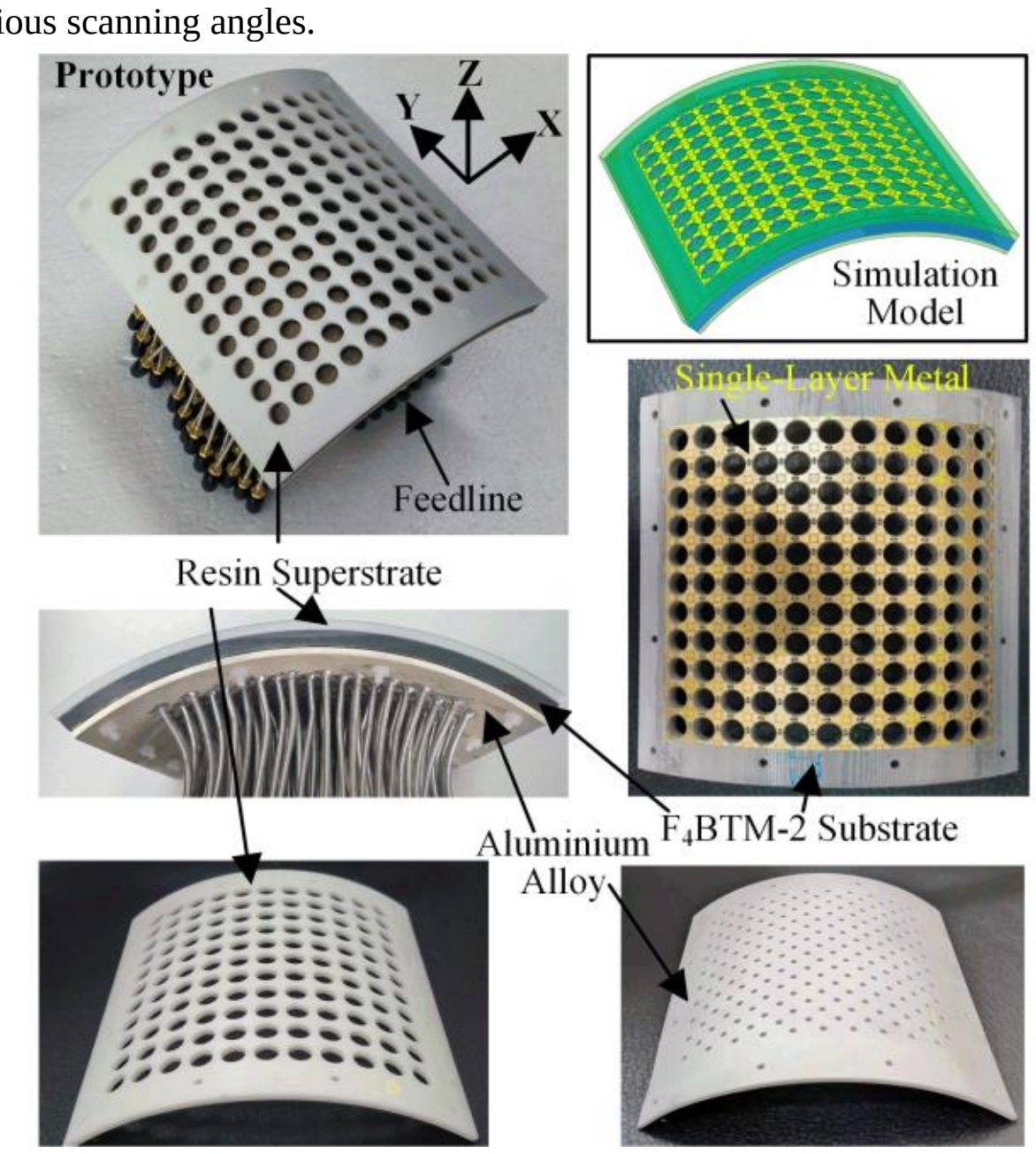

Fig. 7. Configuration and prototype photographs of the proposed 8×8 UDCA.

To ensure the performance stability of the antenna under conformal bending, a modified tightly coupled array with a single-layer metal structure is designed and optimized by simplifying reported tightly coupled arrays with multi-layer metal radiation structures. The modified tightly coupled array change the coupling way between different-layer metal into the slot coupling of single-layer metal. The detailed structure of the geometry for the presented periodic unit cell is shown in Fig. 5. The single-layer metal radiating structure of a unit cell consisting of dual-polarized dipoles and a capacitive coupling patch, is put onto the top side of a $F_4$BTM-2 substrate ($\varepsilon_r$ = 2.2). The impedance matching of ultra-wideband operation for the dual-polarized dipoles is improved by the circular capacitive coupling patch. The feed structure of each dipole consists of an inner conductor of a coaxial line and a grounded metal vias. Such a simple feed structure provides excellent conditions for large-curvature conformal design. To further improve the impedance matching and large-angle beam scanning capability, the resin superstrate ($\varepsilon_r$ = 3.5) is used as a wide-angle impedance matching (WAIM) layer. Meanwhile, the WAIM layer and $F_4$BTM-2 substrate are cylindrically cut out to ensure stable impedance performance under a wide-scanning angle. The metal ground plate serves to connect the outer conductors of the coaxial line and as a mechanical support for the proposed array. The topology of the overall proposed planar array can be regarded as a straightforward three-layer substrate consisting of, from top to bottom, a resin superstrate layer, a substrate electroplated layer, and a metal ground.

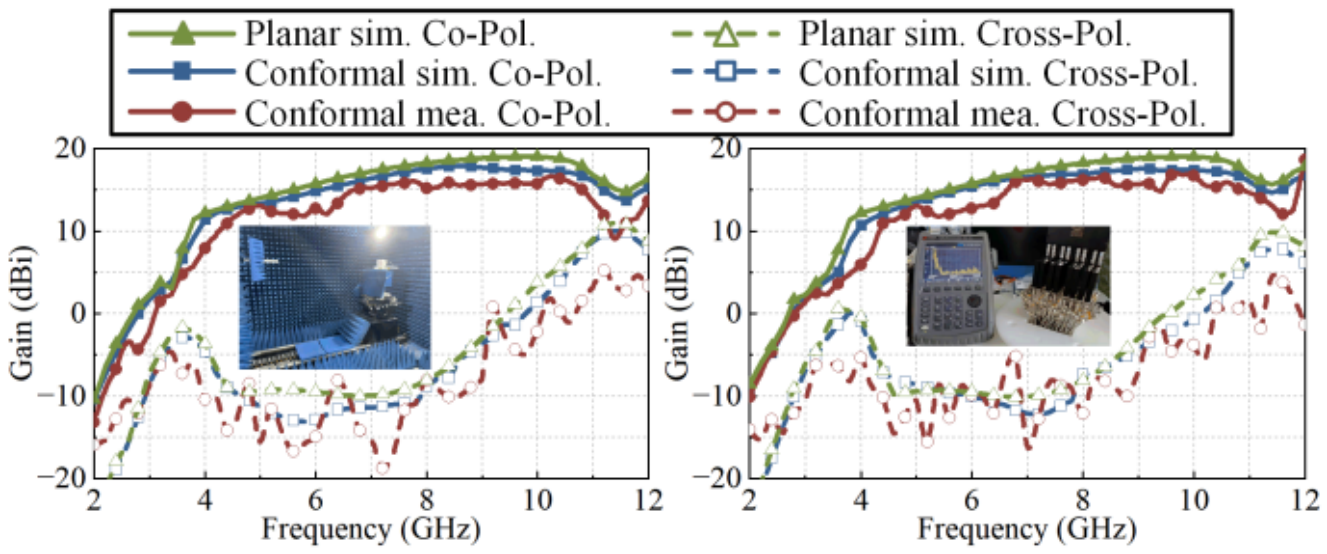

Fig. 8. Calculated and simulated antenna gains in broadside for the 8×8 planar array and UDCA. (a) X-polarized ports operation. (b) Y-polarized ports operation.

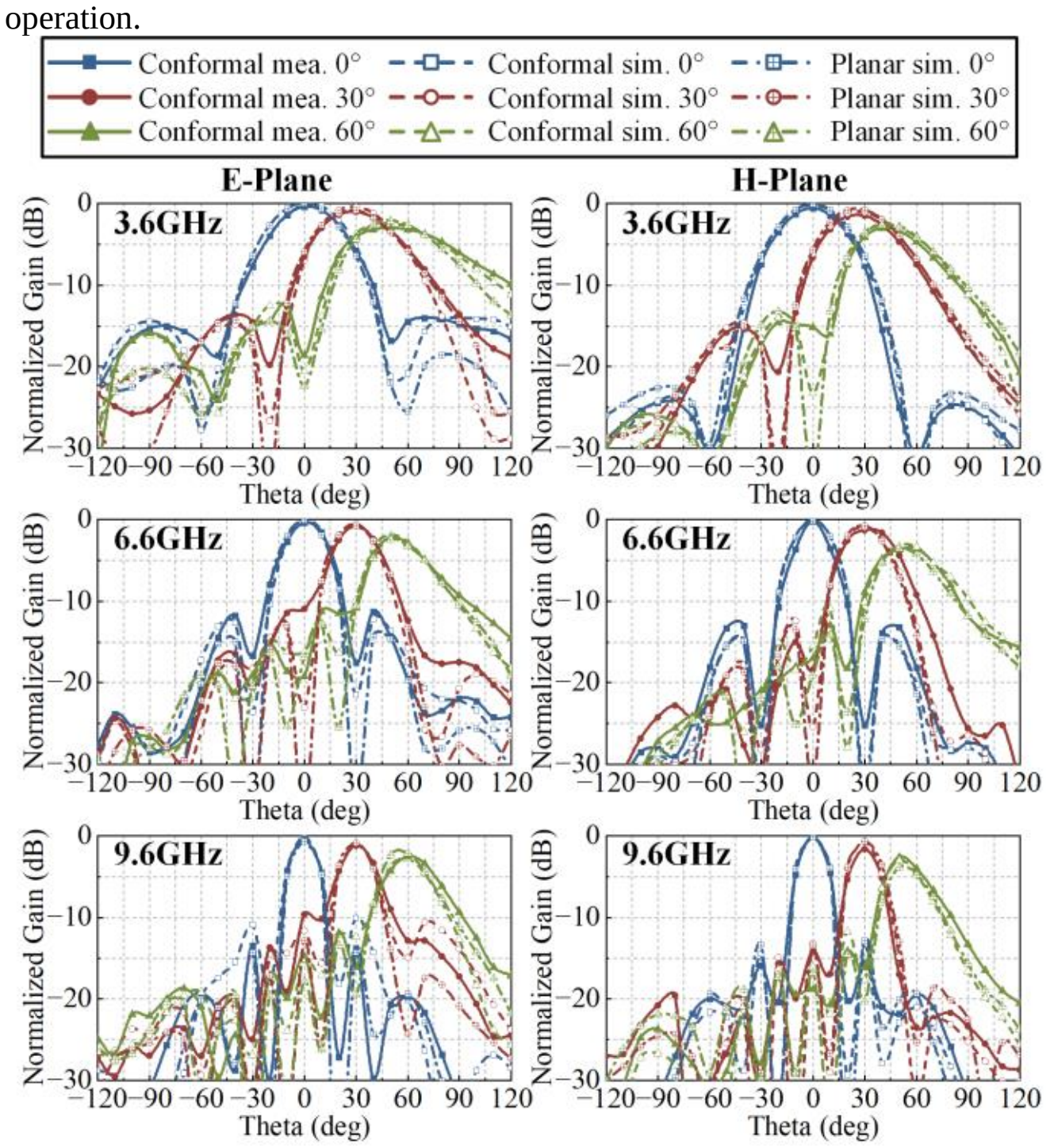

Fig. 9. Calculated and simulated normalized radiation patterns for the 8×8 planar array and UDCA with X-polarization operation.

The simulated active VSWRs of X-polarized unit cell for an infinite array are shown in Fig. 6. Within ±60° scanning of E- and H-planes, the operated band of active VSWR < 3 is 3.5 ~ 9.7 GHz, which means that the infinite array can achieve ±60° wide-scanning angles in the above band.

## IV. Conformal Finite Array Implementation

As shown in Fig. 7, an 8 × 8 conformal array is formed along a cylindrical surface with a 100-mm curvature radius, and an 8×8 UDCA prototype is fabricated, assembled, and measured. It consists of resin superstrate, $F_4$BTM-2 substrate, and aluminium alloy pressed together and fed through 200 feeding coaxial lines, which secures the three layers utilizing plastic studs running vertically. A cube-shaped $F_4$BTM-2 substrate is milled into the required cylindrical structure. The cylindrically conformal $F_4$BTM-2 substrate is laser punched to ensure the accuracy of the curved surface processing, and the plating method is used to plate copper on the $F_4$BTM-2 substrate.

The far-field parameter measurements are calculated by using the unit pattern synthesis [36]. The calculated and simulated antenna gains in broadside for the 8×8 planar array and UDCA are shown in Fig. 8. For the operating band of 3.6~9.6 GHz for the UDCA, the maximum gain is 16.5 dB with cross-polarization better than 10.1 dB. The calculated and simulated normalized X-polarized radiation patterns for the proposed 8×8 UDCA prototype are displayed in Fig. 9. For X-polarized scanning, the UDCA prototype has a gain drop of 2.9/4.0 dB within a 60° E-/H-plane scanning range. Meanwhile, the simulation results for planar arrays are highly similar to those for conformal arrays, verifying the correctness of CMA.

TABLE I
Comparison of The Related Recently Tightly Coupled Arrays

| Ref. | BW* | Element Size (λ) | Scan Range | Pol.* | Curvature Radius |
|---|---|---|---|---|---|
| [24] | 110% | 0.13×0.13×0.07 | ±45° ±45° | Dual | NA* |
| [27] | 122% | 0.11×0.11×0.14 | ±45° ±45° | Dual | NA* |
| [29] | 66.7% | 0.24×0.24×0.10 | ±60° ±60° | Single | 250 mm (5λ) |
| [30] | 100% | 0.15×0.15×0.14 | ±60° ±60° | Single | 150 mm (3.0λ) |
| [31] | 160% | 0.06×0.06×0.05 | Broadside | Single | 200 mm (1.33λ) |
| **Pro.** | **90.9%** | **0.13×0.13×0.09** | **±60° ±60°** | **Dual** | **100 mm (1.2λ)** |

BW*: bandwidth. Pol*: polarization. NA*: not available for planar antenna.
λ* is the wavelength of the lowest operating frequency in free space.

To highlight the merits of the proposed 8×8 UDCA, Table I compares the comprehensive performances between the recently published tightly coupled arrays. Conventional tightly coupled arrays possess excellent characteristics of ultra-wideband and wide-angle scanning, but the complex structures make it difficult to evaluate internal coupling relationships in conformal development, which is used for planar array applications [24], [27]. The cylindrical-conformal tightly coupled arrays are proposed with ultrawideband bandwidth and ±60° wide-scanning characteristics, where they realize only single-polarized operation [29] [30]. Although a flexible tightly coupled array for 9:1 operating bandwidth has been achieved with a suitable 1.3λ radius of curvature, however, its single-polarized operation and lack of beam scanning capability limit its practical application [31]. The proposed UDCA prototype obtains 90.9% relative operating bandwidth with 1.2λ (100 mm) curvature radius and has ± 60° wide-scanning capability in the E- and H-plane, which achieves a balanced high operating performance.

## V. Conclusion

This work contributes to the advancement of conformal phase array technology, offering a feasible and efficient solution for applications requiring low-cost, high-accurate, dual-polarized, ultra-wideband, and wide-angle scanning performance in large-curvature conformal antenna design. The insights obtained from the CMA provide valuable guidance for the development in conformal antennas, with potential applications in various communication and radar systems.